\def\year{2020}\relax
\documentclass[letterpaper]{article} 
\usepackage{aaai20}  
\usepackage{times}  
\usepackage{helvet} 
\usepackage{courier}  
\usepackage[hyphens]{url}  
\usepackage{graphicx} 
\usepackage{graphicx}  
\usepackage{dsfont}
\usepackage{multirow, makecell}
\newcommand\Tstrut{\rule{0pt}{2.6ex}}         
\newcommand\Bstrut{\rule[-1.3ex]{0pt}{0pt}}   
\usepackage{enumitem}
\usepackage{bbold} 
\usepackage{hhline}
\usepackage{amsmath, amssymb, array, amsfonts}
\newcolumntype{L}[1]{>{\raggedright\let\newline\\\arraybackslash\hspace{0pt}}m{#1}}
\newcolumntype{C}[1]{>{\centering\let\newline\\\arraybackslash\hspace{0pt}}m{#1}}
\newcolumntype{R}[1]{>{\raggedleft\let\newline\\\arraybackslash\hspace{0pt}}m{#1}}

\title{M3ER: Multiplicative Multimodal Emotion Recognition using Facial, Textual, and Speech Cues}
\author{Trisha Mittal\textsuperscript{\rm 1}, Uttaran Bhattacharya\textsuperscript{\rm 1}, Rohan Chandra\textsuperscript{\rm 1}, \Large \textbf{Aniket Bera\textsuperscript{\rm 1}, Dinesh Manocha\textsuperscript{\rm 1}}\\ 
\textsuperscript{\rm 1}Department of Computer Science, University of Maryland, College Park, USA\\ 
\{trisha, uttaranb, rohan, ab, dm\}@cs.umd.edu\\ 
Project URL: https://gamma.umd.edu/m3er
}
\begin{document}
\maketitle
\begin{abstract}
We present M3ER, a learning-based method for emotion recognition from multiple input modalities. Our approach combines cues from multiple co-occurring modalities (such as face, text, and speech) and also is more robust than other methods to sensor noise in any of the individual modalities. M3ER models a novel, data-driven multiplicative fusion method to combine the modalities, which learn to emphasize the more reliable cues and suppress others on a per-sample basis. By introducing a check step which uses Canonical Correlational Analysis to differentiate between ineffective and effective modalities, M3ER is robust to sensor noise. M3ER also generates proxy features in place of the ineffectual modalities. We demonstrate the efficiency of our network through experimentation on two benchmark datasets, IEMOCAP and CMU-MOSEI. We report a mean accuracy of $82.7\%$ on IEMOCAP and $89.0\%$ on CMU-MOSEI, which, collectively, is an improvement of about $5\%$ over prior work.
\end{abstract}
\section{Introduction}
\label{sec:intro}
The perception of human emotions plays a vital role in our everyday lives. People modify their responses and behaviors based on their perception of the emotions of those around them. For example, one might cautiously approach a person they perceive to be angry, whereas they might be more forthcoming when approaching a person they perceive to be happy and calm. Given the importance of emotion perception, emotion recognition from sensor data is important for various applications, including human-computer interaction~\cite{ERinHCI}, surveillance~\cite{ERinSurveillance1}, robotics, games and entertainment, and more. In this work, we address the problem of perceived emotion recognition rather than recognition of the actual emotional state. 

\begin{figure}[t]
    \centering
    \includegraphics[width=\columnwidth]{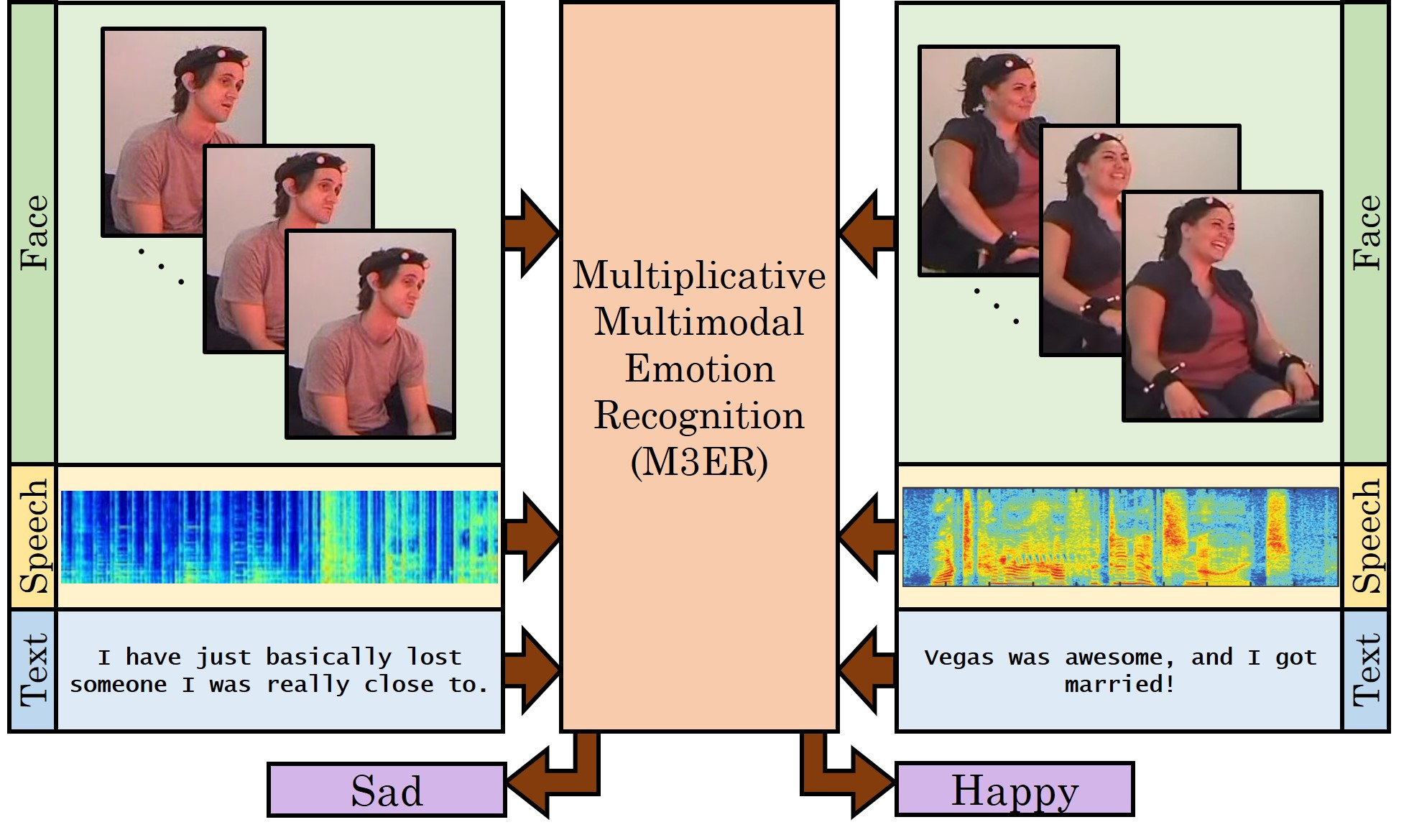}
    \caption{\small{\textbf{Multimodal Perceived Emotion Recognition: }We use multiple modalities to perform perceived emotion prediction. Our approach uses a deep learning model along with a multiplicative fusion method for emotion recognition. We show results on two datasets, IEMOCAP and CMU-MOSEI both of which have face, speech and text as the three input modalities. Above is one sample point extracted from the IEMOCAP dataset. 
    }}
    \label{fig:cover}
\end{figure}

One of the primary tasks in developing efficient AI systems for perceiving emotions is to combine and collate information from the various modalities by which humans express emotion. These modalities include, but are not limited to, facial expressions, speech and voice modulations, written text, body postures, gestures, and walking styles. Many researchers have advocated combining more than one modality to infer perceived emotion for various reasons, including:
\begin{enumerate}[label=(\alph*), noitemsep]
    \item \textit{Richer information: } Cues from different modalities can augment or complement each other, and hence lead to more sophisticated inference algorithms.
    \item \textit{Robustness to Sensor Noise: } Information on different modalities captured through sensors can often be corrupted due to signal noise, or be missing altogether when the particular modality is not expressed, or cannot be captured due to occlusion, sensor artifacts, etc. We call such modalities \textit{ineffectual}. Ineffectual modalities are especially prevalent in in-the-wild datasets. 
\end{enumerate}

However, multimodal emotion recognition comes with its own challenges. At the outset, it is important to decide which modalities should be combined and how. Some modalities are more likely to co-occur than others, and therefore are easier to collect and utilize together. For example, some of the most popular benchmark datasets on multiple modalities, such as IEMOCAP~\cite{iemocap} and CMU-MOSEI~\cite{cmu-mosei}, contain commonly co-occurring modalities of facial expressions with associated speech and transcribed text. With the growing number of social media sites and data on internet~(e.g., YouTube), often equipped with automatic caption generation, it is easier to get data for these three modalities. Many of the other existing multimodal datasets~\cite{recola,afew} are also a subset of these three modalities. Consequently, these are the modalities we have used in our work.

Another challenge is the current lack of agreement on the most efficient mechanism for combining~(also called ``fusing'') multiple modalities~\cite{multimodalmachinelearning}. The most commonly used techniques are early fusion (also ``feature-level'' fusion) and late fusion~(also ``decision-level'' fusion). Early fusion combines the input modalities into a single feature vector on which a prediction is made. In late fusion methods, each of the input modalities is used to make an individual prediction, which is then combined for the final classification. Most prior works on emotion recognition works have explored early fusion~\cite{NDL1} and late fusion~\cite{NDL2} techniques in additive combinations. Additive combinations assume that every modality is always potentially useful and hence should be used in the joint representation. This assumption makes the additive combination not ideal for in-the-wild datasets which are prone to sensor noise. Hence, in our work, we use multiplicative combination, which does not make such an assumption. Multiplicative methods explicitly model the relative reliability of each modality on a per-sample basis, such that reliable modalities are given higher weight in the joint prediction.

\textbf{Main Contributions: } We make the following contributions:
\begin{enumerate}[noitemsep]
    \item We present a multimodal emotion recognition algorithm called M3ER, which uses a data-driven multiplicative fusion technique with deep neural networks. Our input consists of the feature vectors for three modalities --- face, speech, and text. 
    \item To make M3ER robust to noise, we propose a novel preprocessing step where we use Canonical Correlational Analysis~(CCA)~\cite{cca} to differentiate between an ineffectual and effectual input modality signal. 
    \item We also present a feature transformation method to generate proxy feature vectors for ineffectual modalities given the true feature vectors for the effective modalities. This enables our network to work even when some modalities are corrupted or missing.
\end{enumerate}
We compare our work with prior methods by testing our performance on two benchmark datasets IEMOCAP and CMU-MOSEI. We report an accuracy of \textbf{$82.7$\%} on the IEMOCAP dataset and \textbf{$89.0$\%} on the CMU-MOSEI dataset, which is a collective {$5$\%} accuracy improvement on the absolute over prior methods. We show ablation experiment results on both datasets, where almost $75$\% of the data has at least one modality corrupted or missing, to demonstrate the importance of our contributions. As per the annotations in the datasets, we classify IEMOCAP into $4$ discrete emotions~({angry, happy, neutral, sad}) and CMU-MOSEI into $6$ discrete emotions~({anger, disgust, fear, happy, sad, surprise}). According to the continuous space representations, emotions are seen as points on a 3D space of arousal, valence, and dominance~\cite{pad}. The discrete emotions are related to the continuous space through an eigen-transform; therefore we can switch between the representations without adding any noise. 
\section{Related Work}
\label{sec:rw}
In this section, we give a brief overview of previous works on unimodal and multimodal emotion recognition, as well as modality combination techniques that have been used in the broader field of multimodal machine learning.
\begin{figure*}[h!]
    \centering
    \scalebox{0.9}{\includegraphics[width=\linewidth]{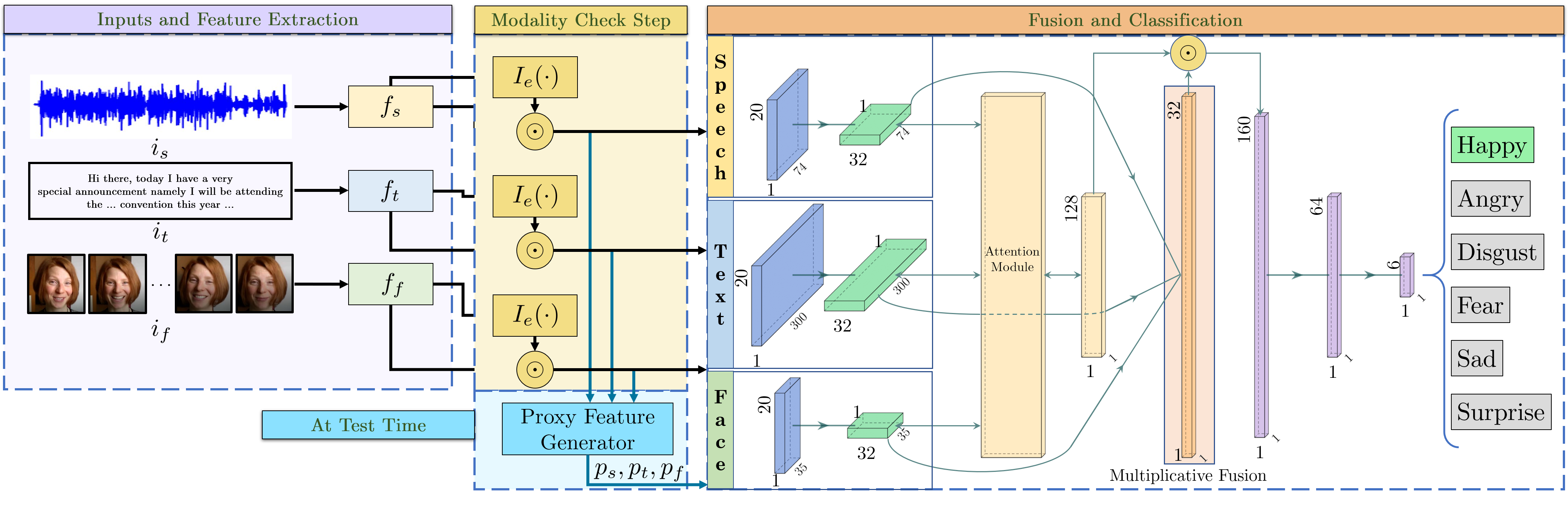}}
    \caption{\small{\textbf{M3ER: }We use three modalities, speech, text and the facial features. We first extract features to obtain $f_{s}$, $f_{t}$, $f_{f}$ from the raw inputs, $i_{s}$, $i_{t}$ and $i_{f}$~(purple box). The feature vectors then are checked if they are effective. We use a indicator function $I_{e}$~(Equation~\ref{eq: indicator}) to process the feature vectors~(yellow box). These vectors are then passed into the classification and fusion network of M3ER to get a prediction of the emotion~(orange box). At the inference time, if we encounter a noisy modality, we regenerate a proxy feature vector ($p_{s}$, $p_{t}$ or $p_{f}$) for that particular modality~(blue box). }}
    \label{fig:overview}
\end{figure*}

\noindent\textbf{Emotion Recognition in Psychology Research: }Understanding and interpreting human emotion is of great interest in psychology. The initial attempts~\cite{psych5} at predicting emotion only from facial expressions were not considered very reflective of the human sensory system and were questioned. There is also unreliability in using facial expressions, because of the ease of displaying ``mocking'' expressions~\cite{psych6}, especially in the presence of an audience~\cite{psych7}. Psychology research also points to the importance of considering cues other than facial expressions to make more accurate predictions. Sebe et al.~\shortcite{psych2}, Aviezer et al.~\shortcite{psych1} and Pantic et al.~\shortcite{psych3} highlight the fact that an ideal system for automatic human emotion recognition should be multimodal, because this is more close to the human sensory system. Meeran et al.~\shortcite{psych4} suggest that the integration of modalities is an inevitable step learned very early-on in the human sensory system. \\

\noindent\textbf{Unimodal Emotion Recognition: }The initial attempts in human emotion recognition have been mostly unimodal. Even in that domain, the most predominantly explored modality has been facial expressions~\cite{face1,face2}, owing to the availability of face datasets and advances in computer vision methods. Other modalities that have been explored include speech or voice expressions~\cite{speech1}, body gestures~\cite{body}, and physiological signals such as respiratory and heart signals~\cite{physiological1}.\\

\noindent \textbf{Multimodal Emotion Recognition: } Multimodal emotion recognition was initially explored using classifiers like Support Vector Machines, and linear and logistic regression~\cite{NDL1,NDL2,NDL3}, when the size of the datasets was less than 500. As bigger datasets were developed, deep learning architectures~\cite{DL1,DL2,DL3,DL4,DL5,DL6} were explored. All multimodal methods also perform feature extraction steps on each of the input modalities, using either hand-crafted formulations or deep learning architectures. Some of the architectures that have been explored are Bi-Directional Long Short Term Memory (BLSTM) networks~\cite{DL1},  Deep Belief Networks (DBNs)~\cite{DL2}, and Convolutional Neural Networks~\cite{DL5}. Other methods are based on hierarchical networks~\cite{DL3} and Relational Tensor Networks~\cite{DL6}. \\

\noindent \textbf{Modality Combination: } Prior works in emotion recognition~\cite{NDL1,NDL2,NDL3,DL1,DL2} using either late or early fusion have relied on additive combinations.  The performance of these additive approaches relies on figuring out the relative emphasis to be placed on different modalities. However, in the real-world, not every modality is equally reliable for every data point due to sensor noise, occlusions, etc. Recent works have also looked at variations on more sophisticated data-driven~\cite{DL5}, hierarchical~\cite{DL3}, and attention-mechanism based~\cite{DL1,DL5} fusion techniques. Multiplicative combination methods~\cite{multiplicative} explicitly models the relative reliability of each modality, such that more reliable modalities are given more weight in the joint prediction. Reliable modalities can also change from sample to sample, so it is also important to learn which modalities are more reliable on a per sample basis. This method has previously been shown to be successful on tasks like user profiling and physical process recognition~\cite{multiplicative}. \\

\noindent \textbf{Canonical Correlational Analysis (CCA): }The objective of CCA~\cite{cca} is to project the input vectors into a common space by maximizing their component-wise correlation. There have been extensions to CCA, namely Deep CCA~\cite{dcca}, Generalized CCA~\cite{gcca}, and Kernel CCA~\cite{kcca}, which learn parametric non-linear transformations of two random vectors, such that their correlation is maximized. CCA approaches have also been explored for the task of multimodal emotion recognition~\cite{ERwithCCA}, to get maximally correlated feature vectors from each input modality before combining them. In our work, we use CCA to check for correlation among the input modalities and to check for effective and ineffective modalities.
\section{M3ER: Our Approach} 
\label{sec:approach}
\begin{figure*}[h!]
    \centering
    \includegraphics[width=\linewidth]{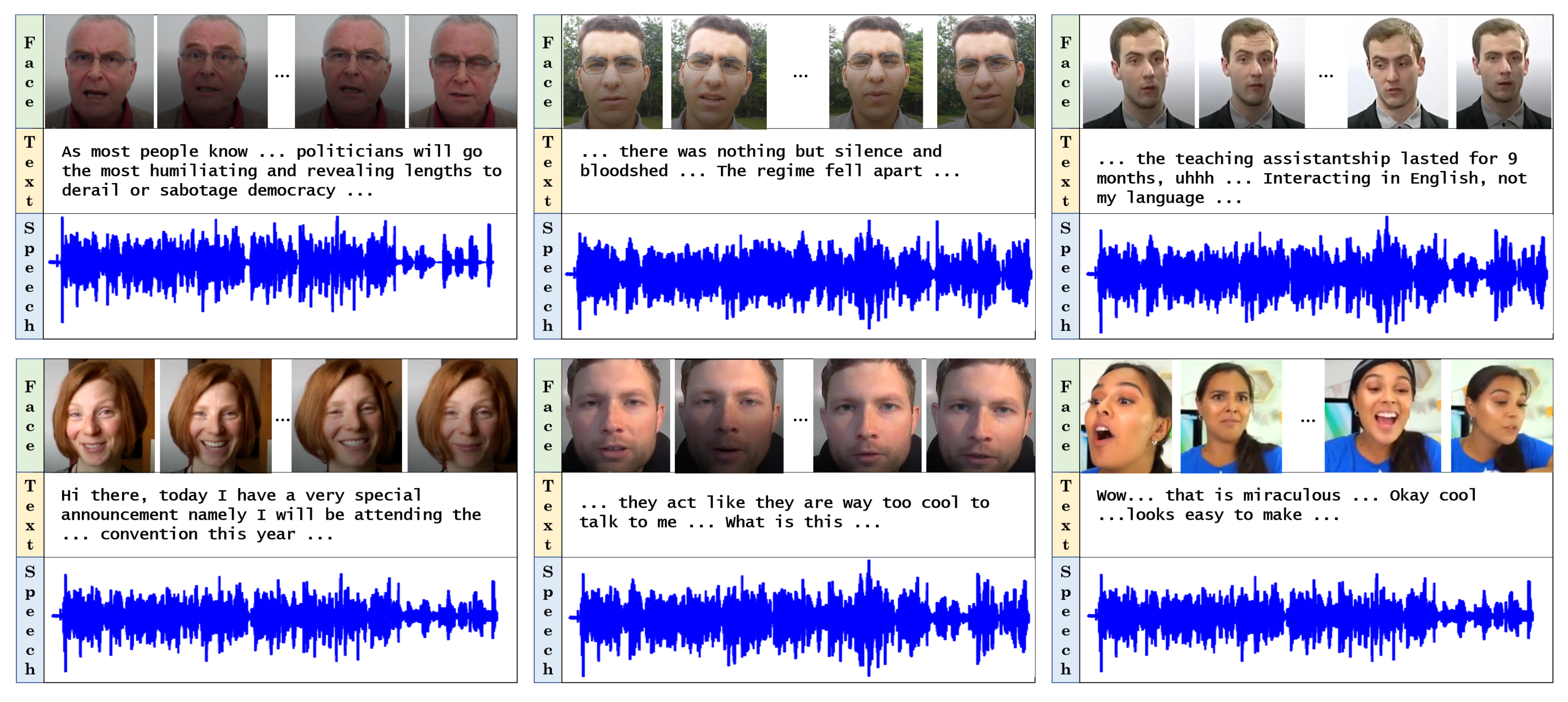}
    \caption{\small{\textbf{Qualitative Results on CMU-MOSEI: }We qualitatively show data points correctly classified by M3ER from all the $6$ class labels of CMU-MOSEI. The labels as classified by M3ER in row order from top left, are \textit{Anger, Disgust, Fear, Happy, Sad, Surprise}.}}
    \label{fig:qualitative}
\end{figure*}
\begin{figure}[h!]
    \centering
    \includegraphics[width=\columnwidth]{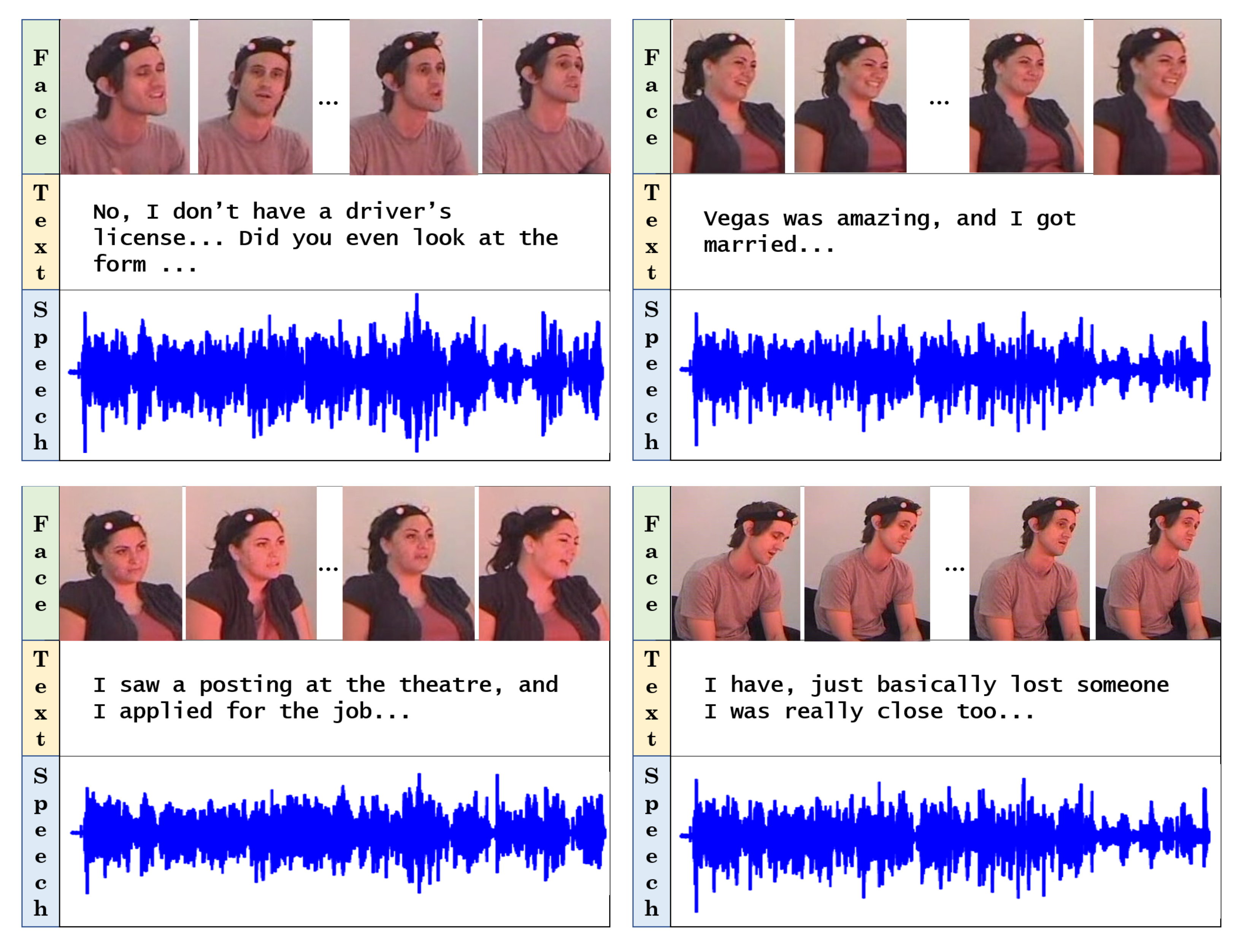}
    \caption{\small{\textbf{Qualitative Results on IEMOCAP: }We qualitatively show data points correctly classified by M3ER from all the $4$ class labels of IEMOCAP. The labels as classified by M3ER in row order from top left, are \textit{Angry, Happy, Neutral, Sad}.}}
    \label{fig:qualitative1}
\end{figure}
\subsection{Notation}
\label{sec:notation}
We denote the set of modalities as $\mathcal{M} = \left \{\textrm{face}, \textrm{text}, \textrm{speech} \right\}$. The feature vectors for each modality are denoted as $f_f$, $f_t$, and $f_s$, respectively. 
We denote the set of predicted emotions as $\mathcal{E} = \{ \textrm{happy}, \textrm{sad}, \textrm{angry}, \textrm{neutral} \}$. The proxy feature vectors generated for speech, text, and face vectors are represented by $p_s, p_t, p_f$, respectively. Finally, we define an indicator function, $I_{e}(f)$ that outputs either a vector of zero or one of the same dimension as $f$, depending on the conditions of the function definition.
\subsection{Overview}
\label{sec:overview}
We present an overview of our multimodal perceived emotion recognition model in Figure \ref{fig:overview}. During training, we first extract feature vectors ($f_s, f_t, f_f$) from raw inputs ($i_s, i_t, i_f$) (purple box in the Figure \ref{fig:overview}). These are then passed through the modality check step (yellow box in the Figure \ref{fig:overview}) to distinguish between effective and ineffectual signals, and discarding the latter if any (See Section~\ref{sec:modalitycheck}). The feature vectors as returned by the modality check step go through three deep-layered feed-forward neural network channels (orange box in Figure \ref{fig:overview}). Finally, we add our multiplicative fusion layer to combine the three modalities.  At test time, the data point once again goes through the modality check step. If a modality is deemed ineffectual, we regenerate a proxy feature vector (blue box in Figure \ref{fig:overview}) which is passed to the network for the emotion classification. 
In the following subsections, we explain each of the three novel components of our network in detail.
\subsection{Modality Check Step}
\label{sec:modalitycheck}
To enable perceived emotion recognition in real world scenarios, where sensor noise is inevitable, we introduce the Modality Check step which filters ineffectual data.
It has been observed in emotion prediction studies~\cite{ERwithCCA}, that for participants whose emotions were predicted correctly, each of their corresponding modality signals correlated with at least one other modality signal. We directly exploit this notion of correlation to distinguish between features that could be effective for emotion classification (effective features) and features that are noisy~(ineffectual features).\\
More concretely, we use Canonical Correlation Analysis (CCA) to compute the correlation score, $\rho$, of every pair of input modalities. 
Given a pair of feature vectors, $f_i, f_j$, with $i,j \in \mathcal{M}$, we first compute the the projective transformations,  $H^{i}_{i,j}$ and $H^{j}_{i,j}$, for both feature vectors, respectively. Also note that these feature vectors $f_i, f_j$ are reduced to the same lower dimensions ($100$, here).  We obtain the projected vector by applying the projective transformation. Thus, in our example above,

\begin{equation*} 
    f^{'}_{i} = H^{i}_{i,j}  f_i,
\end{equation*} 
and, 
\begin{equation*} 
    f^{'}_{j} = H^{j}_{i,j}  f_j,
\end{equation*} 
\noindent Finally, we can compute the correlation score for the pair $\{f_i, f_j\}$ using the formula:
\[ \rho(f^{'}_i, f^{'}_j) = \frac{\textrm{cov}(f^{'}_i, f^{'}_j)}{\sigma_{f^{'}_i}  \sigma_{f^{'}_j}}\]
and check them against an empirically chosen threshold $(\tau)$.
$\forall i \in m$, we check
\begin{equation*}
    \rho(f^{'}_i, f^{'}_j) < \tau, 
\end{equation*} 
\noindent where $\forall \left ( i,j \right ) \in \mathcal{M}, i \neq j$.

For implementation purposes, we keep the $H^{j}_{i,j}$ for all pairs of modalities precomputed based on the training set. At inference time, we simply compute the projected vectors $f^{'}_{i},f^{'}_{j}$ and $\rho(f^{'}_i, f^{'}_j)$.

We compare the correlation against a heuristically chosen threshold, $\tau$ and introduce the following indicator function,
\begin{equation}
\label{eq: indicator}
I_{e}(f_i) = 
\begin{cases}
\mathbb{0}& \rho(f_i,f_j) < \tau, (i,j) \in \mathcal{M}, i \neq j, \\ 
\mathds{1} & else.
\end{cases}
\end{equation}
\noindent For all features, we apply the following operation, $I_{e}(f) \odot f$, which discards ineffectual features and retains the effective ones. Here, $\odot$ denotes element-wise multiplication. 
\subsection{Regenerating Proxy Feature Vectors} 
When one or more modalities have been deemed ineffectual at test time in the modality check step, we generate proxy feature vectors for the ineffectual modalities using the following equation, $p_{{i}} = \mathcal{T}f_{{i}}$, where $i \in \mathcal{M}$ and $\mathcal{T}$ is any linear transformation. We illustrate the details below.

Generating exact feature vectors for missing modalities is challenging due to the non-linear relationship between the modalities. However, we empirically show that by relaxing the non-linear constraint, there exists a linear algorithm that approximates the feature vectors for the missing modalities with high classification accuracy. We call these resulting vectors: proxy feature vectors.\\
Suppose that during test time, the feature vector for the speech modality is corrupt and identified as ineffectual, while $f_f$ is identified as effective during the Modality Check Step. Our aim is then to regenerate a proxy feature vector, $p_{s}$, for the speech modality. More formally, we are given, say, a new, unseen face modality feature vector, $f_{{f}}$, the set of observed face modality vectors, $\mathcal{F} = \{  f_1,f_2, \ldots , f_n \}$, and the set of corresponding observed speech modality vectors, $\mathcal{S} = \{  s_1,s_2, \ldots , s_n \}$. Our goal is to generate a proxy speech vector, $p_{{s}}$, corresponding to $f_{{f}}$. 
We begin by preprocessing the inputs to construct bases, $\mathcal{F}_b = \{  v_1,v_2, \ldots , v_p \}$ and $\mathcal{S}_b = \{  w_1,w_2, \ldots , w_q \}$ from the column spaces of $\mathcal{F}$ and $\mathcal{S}$. Under the relaxed constraint, we assume there exists a linear transformation, $\mathcal{T}:\mathcal{F}_b\rightarrow \mathcal{S}_b$. Our algorithm proceeds without assuming knowledge of $\mathcal{T}$:
\begin{enumerate}
    \item The first step is to find $v_j = \textrm{argmin}_j d(v_j, f_f)$, where $d$ is any distance metric. We chose the $L_2$ norm in our experiment
    s. We can solve this optimization problem using any distance metric minimization algorithm such as the K-nearest neighbors algorithm.
    \item Compute constants ${a_i \in \mathbb{R}}$ by solving the following linear system, $f_{{f}} =  \sum_{i=1}^{p}a_i v_i$. Then,
    \begin{equation*}
        \begin{split}
            p_{{s}} = \mathcal{T}f_{{f}} = \sum_{i = 1}^{p} a_i \mathcal{T}v_i = \sum_{i= 1}^{p} a_i w_i.
        \end{split}
    \end{equation*}
    
\end{enumerate}
Our algorithm can be extended to generate proxy vectors from effective feature vectors corresponding to multiple modalities. In this case, we would apply the steps above to each of the effective feature vectors and take the mean of both the resulting proxy vectors. 
\subsection{Multiplicative Modality Fusion}
The key idea in the original work~\cite{multiplicative} for multiplicative combination is to explicitly suppress the weaker (not so expressive) modalities, which indirectly boost the stronger (expressive) modalities. They define the loss for the $i^{th}$ modality as follows.
\begin{equation}
\label{eq:original_loss}
\begin{split}
    c^{(y)} = -\sum_{i=1}^{M}\prod_{j\neq i} \left ( 1 - p_{j}^{(y)} \right )^{\beta/(M-1)}\log p_{i}^{(y)}
\end{split}
\end{equation}
where $y$ is the true class label, $M$ is the number of modalities, $\beta$ is the hyperparameter that down-weights the unreliable modalities and $p_{i}^{(y)}$ is the prediction for class $y$ given by the network for the $i^{th}$ modality. This indirectly boosts the stronger modalities. In our approach, we reverse this concept and propose a modified loss. We explicitly boost the stronger modalities in the combination network. The difference is subtle but has key significance on the results. In the original formulation, the modified loss was given by Equation \ref{eq:original_loss}. We empirically show that the modified loss gives better classification accuracies than the originally proposed loss function in Section \ref{sec:results}. The original loss function tries to ignore or tolerate the mistakes of the modalities making wrong predictions by explicitly suppressing them, whereas in our modified version, we ignore the wrong predictions by simply not addressing them and rather focusing on modalities giving the right prediction. In the original loss, calculating the loss for each modality depends on the probability given by all the other modalities. This has a higher computation cost due to the product term. Furthermore, if either of the input modalities produces an outlier prediction due to noise in the signal, it affects the prediction of all other modalities. Our proposed modified loss is as follows:
\begin{equation}
    \label{eq:modified_loss}
    c^{(y)} = -\sum_{i=1}^{M}          \left ( p_{i}^{(y)} \right )^{\beta/(M-1)}\log p_{i}^{(y)}
\end{equation}
This fusion layer is applied to combine the three input modalities.\\

M3ER is a modular algorithm that can work on top of existing networks for multimodal classification. Given a network for multiple modalities, we can replace the fusion step and incorporate the modality check and proxy vector regeneration of the M3ER and improve classification accuracies. In the next Section, we demonstrate this point by incorporating M3ER in SOTA networks for two datasets, IEMOCAP and CMU-MOSEI.
\begin{table}[t]
\centering
\resizebox{.7\columnwidth}{!}{%
\begin{tabular}{|c|c|c|c|} 
\hline
\textbf{Dataset \Tstrut} & \textbf{Method} & \textbf{F1} & \textbf{MA}\\
\hhline{|=|=|=|=|}
\multirow{5}{*}{IEMOCAP }
& Kim et al. \Tstrut \shortcite{DL2}    & - & 72.8\% \\
& Majumdar et al. \shortcite{DL3}       & - & 76.5\% \\
& Yoon et al. \shortcite{DL1}           & - & 77.6\% \\
\cline{2-4}
& \textbf{M3ER \Tstrut}           & \textbf{0.824} & \textbf{82.7\%} \\
\hhline{|=|=|=|=|}
\multirow{5}{*}{CMU-MOSEI}
& Sahay et al. \Tstrut \shortcite{DL6}  & 0.668 & - \\
& Zadeh et al. \shortcite{DL4}          & 0.763 & - \\
& Choi et al. \shortcite{DL5}           & 0.895 & - \\
\cline{2-4}
& \textbf{M3ER \Tstrut}           & \textbf{0.902} & \textbf{89.0\%} \\
\hline
\end{tabular}
}
\caption{\small{\textbf{M3ER for Emotion Recognition: } We compare the F1 scores and the mean classification accuracies (MA) of M3ER on the two datasets, IEMOCAP and CMU-MOSEI, with three prior SOTA methods. Numbers not reported by prior methods are marked with `-'. We observe around $5$-$10\%$ increase in MA and $1$-$23\%$ increase in F1 score.}}
\label{tab:accuracy}
\end{table}
\section{Implementation Details} 
\label{sec:implementation}
We state the implementation and training details for training with M3ER on the CMU-MOSEI dataset in this section. Details on the network, implementation, and training on the IEMOCAP dataset can be found here \footnote{\url{https://github.com/TrishaMittal/M3ER}}).
\subsection{Feature Extraction}
To extract $f_t$ from the CMU-MOSEI dataset, we use the 300-dimensional pre-trained GloVe word embeddings~\cite{glove}. To compute $f_s$ from the CMU-MOSEI dataset, we follow the approach of Zadeh et al.~\shortcite{DL4} and obtain the 12 Mel-frequency cepstral coefficients, pitch, voiced/unvoiced segmenting features, glottal source parameters among others. Lastly, to obtain $f_f$, we use the combination of face embeddings obtained from state-of-the-art facial recognition models, facial action units, and facial landmarks for CMU-MOSEI.
\subsection{Classification Network Architecture} 
\label{subsec:classificationarchitecture}
For training on the CU-MOSEI dataset, we integrate our multiplicative fusion layer into Zadeh et al.'s~\shortcite{cmu-mosei-arch} memory fusion network~(MFN). Each of the input modalities is first passed through single-hidden-layer LSTMs, each of output dimension $32$. The outputs of the LSTMs, along with a 128-dimensional memory variable initialized to all zeros (yellow box in the network Figure \ref{fig:overview}), are then passed into an~\textit{attention module} as described by the authors of MFN. The operations inside the attention module are repeated for a fixed number of iterations $t$, determined by the maximum sequence length among the input modalities~($t=20$ in our case). The outputs at the end of every iteration in the attention module are used to update the memory variable as well as the inputs to the LSTMs. After the end of $t$ iterations, the outputs of the $3$ LSTMs are combined using multiplicative fusion to a $32$ dimensional feature vector. This feature vector is concatenated with the final value of the memory variable, and the resultant $160$ dimensional feature vector is passed through a $64$ dimensional fully connected layer followed by a $6$ dimensional fully connected to generate the network outputs.
\subsection{Training Details}
\label{subsec:training}
For training with M3ER on the CMU-MOSEI dataset, we split the CMU-MOSEI dataset into training~($70\%$), validation~($10\%$), and testing~($20\%$) sets. We use a batch size of $256$ and train it for $500$ epochs. We use the Adam optimizer~\cite{adam} with a learning rate of $0.01$. All our results were generated on an NVIDIA GeForce GTX 1080 Ti GPU. 
\begin{figure}[t]
\includegraphics[width=\columnwidth]{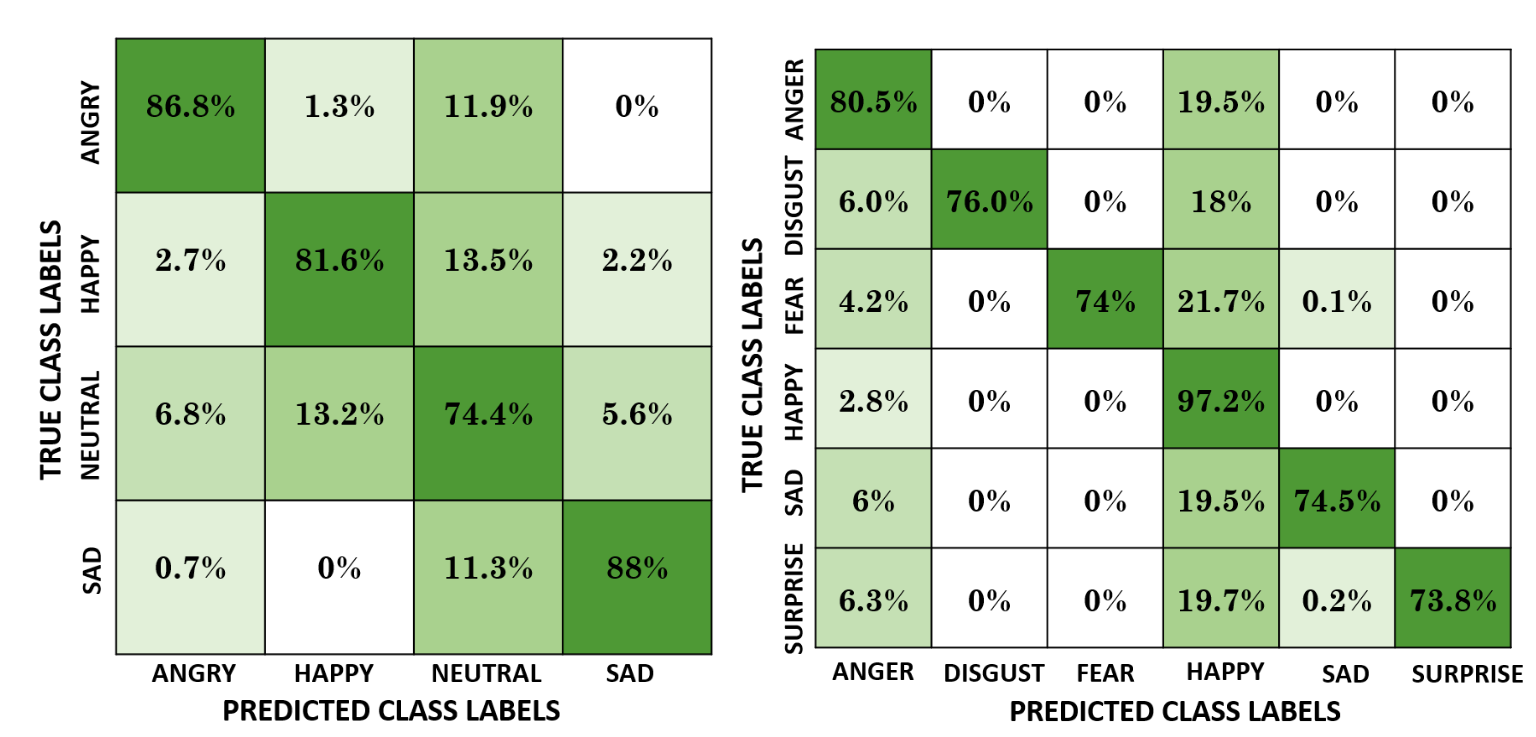}
\caption{\small{\textbf{Confusion Matrix: }For each emotion class, we show the percentage of inputs belonging to that class that were correctly classified by M3ER (dark green cells) and the percentage of inputs that were misclassified into other classes (pale green and white cells) for both the datasets. \textit{Left:} Confusion matrix for classification on IEMOCAP dataset. \textit{Right:} Confusion matrix for classification on CMU-MOSEI dataset.}}
\label{fig:confusionmatrix}
\end{figure}
\section{Experiments and Results}
\label{sec:results}
We perform experiments on the two large-scale benchmark datasets, IEMOCAP and CMU-MOSEI, described in Section \ref{subsec:dataset}. In Section~\ref{subsec:classification}, we list the SOTA algorithms with which we compare M3ER using standard classification evaluation metrics. We report our findings and analysis in Section~\ref{subsec:analysis}. We perform exhaustive ablation experiments to motivate the benefits of our contributions in Section~\ref{subsec:ablation}. Finally, we provide details of all hyperparameters and the hardware used for training M3ER in Section \ref{subsec:training}.
\subsection{Datasets}
\label{subsec:dataset}
The \textit{Interactive Emotional Dyadic Motion Capture}~(IEMOCAP) dataset~\cite{iemocap} consists of text, speech, and face modalities of $10$ actors recorded in the form of conversations using a Motion Capture camera. The conversations include both scripted and spontaneous sessions. The labeled annotations consists of four emotions --- angry, happy, neutral, and sad. The \textit{CMU Multimodal Opinion Sentiment and Emotion Intensity}~(CMU-MOSEI)~\cite{cmu-mosei} contains $23,453$ annotated video segments from $1,000$ distinct speakers and $250$ topics acquired from social media channels. The labels in this dataset comprise six emotions --- angry, disgust, fear, happy, sad and surprise.
\begin{figure}[h]
\centering
\scalebox{0.95}{\includegraphics[width=\columnwidth]{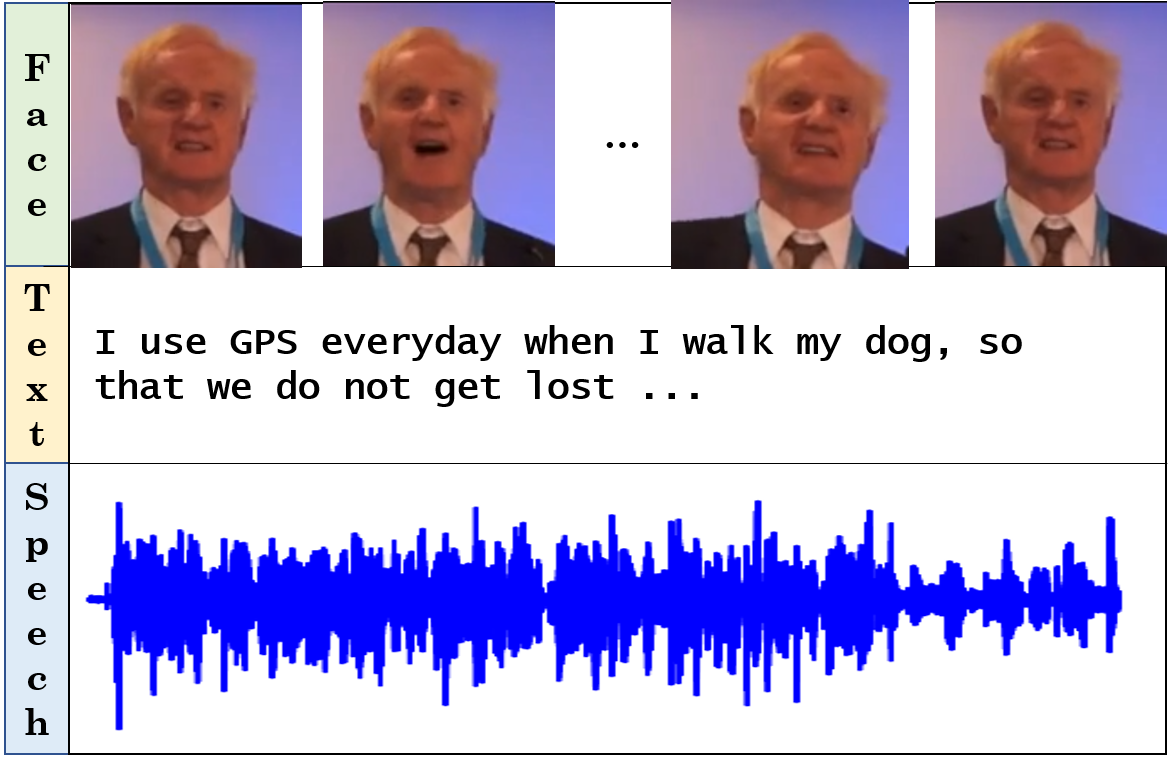}}
\caption{\small{\textbf{Misclassification by M3ER: }This is the text and face input of a `happy' data point from CMU-MOSEI dataset that our model, M3ER misclassifies as `angry'. Here, the man is giving a funny speech with animated and exaggerated facial looks which appear informative but lead us to a wrong class label.}}
\label{fig:failure}
\end{figure}
\begin{table*}[h!]
\centering
(a) Ablation Experiments performed on IEMOCAP Dataset.\\
\resizebox{.97\linewidth}{!}{%
\begin{tabular}{|c|c|c|c|c|c|c|c|c|c|c|c|}
\hline
\textbf{Ineffectual}                                    &
\multirow{2}{*}{\textbf{Experiments} }                  &
\multicolumn{2}{c|}{\textbf{Angry \Tstrut }}            &
\multicolumn{2}{c|}{\textbf{Happy }}                    &
\multicolumn{2}{c|}{\textbf{Neutral }}                  &
\multicolumn{2}{c|}{\textbf{Sad }}                      &
\multicolumn{2}{c|}{\textbf{Overall }}
\\\cline{3-12}
\textbf{modalities?}&&
\textbf{F1 \Tstrut} & \textbf{MA \Bstrut} &
\textbf{F1} & \textbf{MA} &
\textbf{F1} & \textbf{MA} &
\textbf{F1} & \textbf{MA} &
\textbf{F1} & \textbf{MA} \\
\hhline{|=|=|=|=|=|=|=|=|=|=|=|=|}
\multirow{2}{*}{No} &
Original Multiplicative Fusion~\cite{multiplicative} \Tstrut &  0.794 &  80.6\% &  0.750 & 76.9\%  & 0.695  & 68.0\%   & 0.762  & 80.8\%  & 0.751  & 76.6\%  \\
\cline{2-12}
& \textbf{M3ER \Tstrut} & \textbf{0.862} & \textbf{86.8\% } & \textbf{0.862 } &  \textbf{81.6\%}  &\textbf{ 0.745}  &\textbf{ 74.4\%}  & \textbf{0.828 } & \textbf{88.1\%}  &  \textbf{0.824}  & \textbf{82.7\% }   \\
\hhline{|=|=|=|=|=|=|=|=|=|=|=|=|}
\multirow{3}{*}{Yes} &
M3ER -- Modality Check Step -- Proxy Feature Vector \Tstrut & 0.704  & 71.6\%  & 0.712  & 70.4\%  & 0.673  & 64.7\%   & 0.736  & 79.8\%  & 0.706  & 71.6\%  \\
& M3ER -- Proxy Feature Vector \Tstrut &  0.742 & 75.7\%  & 0.745  & 73.7\%  & 0.697  & 66.9\%   & 0.778  & 84.0\%  & 0.741  & 75.1\%  \\
\cline{2-12}
& \textbf{M3ER \Tstrut}  &  \textbf{0.799} & \textbf{82.2\%}  & \textbf{0.743}  & \textbf{76.7\%}  & \textbf{0.727}  & \textbf{67.5\%}   & \textbf{0.775}  & \textbf{86.3\%}  &  \textbf{0.761} & \textbf{78.2\%}  \\
\hline
\end{tabular}
}

\bigskip 
\centering
(b) Ablation Experiments performed on CMU-MOSEI Dataset.\\
\resizebox{0.97\linewidth}{!}{%
\begin{tabular}{|c|c|c|c|c|c|c|c|c|c|c|c|c|c|c|c|}
\hline
\textbf{Ineffectual}                                    &
\multirow{2}{*}{\textbf{Experiments} }                  &
\multicolumn{2}{c|}{\textbf{Angry \Tstrut }}            &
\multicolumn{2}{c|}{\textbf{Disgust }}                  &
\multicolumn{2}{c|}{\textbf{Fear }}                     &
\multicolumn{2}{c|}{\textbf{Happy }}                    &
\multicolumn{2}{c|}{\textbf{Sad }}                      &
\multicolumn{2}{c|}{\textbf{Surprise }}                 &
\multicolumn{2}{c|}{\textbf{Overall }}
\\\cline{3-16}
\textbf{modalities?}&&
\textbf{F1 \Tstrut} & \textbf{MA \Bstrut} &
\textbf{F1} & \textbf{MA} &
\textbf{F1} & \textbf{MA} &
\textbf{F1} & \textbf{MA} &
\textbf{F1} & \textbf{MA} &
\textbf{F1} & \textbf{MA} &
\textbf{F1} & \textbf{MA} \\
\hhline{|=|=|=|=|=|=|=|=|=|=|=|=|=|=|=|=|}
\multirow{2}{*}{No} &
Original Multiplicative Fusion~\cite{multiplicative} \Tstrut & 0.889  & 79.9\%  & 0.945 & 89.6\%  & 0.963  & 93.1\%  & 0.587  & 55.8\%  &  0.926  & 85.3\%  & 0.949  & 90.0\% & 0.878 & 82.3\% \\
\cline{2-16}
& \textbf{M3ER \Tstrut} & \textbf{0.919} & \textbf{86.3\%}  & \textbf{0.927}  &\textbf{ 92.1\% } & \textbf{0.904} & \textbf{88.9\%}  & \textbf{0.836}  & \textbf{82.1\%}  & \textbf{0.899}   & \textbf{89.8\%}  & \textbf{0.952}  &\textbf{ 95.0\%} & \textbf{0.902} & \textbf{89.0\%} \\
\hhline{|=|=|=|=|=|=|=|=|=|=|=|=|=|=|=|=|}
\multirow{3}{*}{Yes} &
M3ER -- Modality Check Step -- Proxy Feature Vector \Tstrut & 0.788 & 73.3\%  & 0.794  & 80.0\%  & 0.843  & 85.0\%  & 0.546   & 55.7\%  & 0.832  & 79.5\%  & 0.795 & 80.1\%  & 0.764 & 75.6\% \\
& M3ER -- Proxy Feature Vector \Tstrut & 0.785  & 77.8\%  & 0.799  & 83.2\%  & 0.734  & 77.5\% &  0.740 & 77.1\%  & 0.840  & 86.0\%  & 0.781  & 83.5\% & 0.783 & 80.9\% \\
\cline{2-16}
& \textbf{M3ER \Tstrut} & \textbf{0.816}  & \textbf{81.3\%}  & \textbf{0.844}  &  \textbf{86.8\%} & \textbf{0.918}  & \textbf{89.4\%}   & \textbf{0.780}  & \textbf{75.7\%}  & \textbf{0.873}  &  \textbf{86.1\%}  & \textbf{0.932}  &  \textbf{91.3\%} & \textbf{0.856} & \textbf{85.0\%} \\
\hline
\end{tabular}
}
\caption{\small{\textbf{Ablation Experiments: }We remove one component of M3ER at a time, and report the F1 and MA scores on the IEMOCAP and the CMU-MOSEI datasets, to showcase the effect of each of these components. Modifying the loss function leads to an increase of $6$-$7\%$ in both F1 and MA. Adding the modality check step on datasets with ineffectual modalities leads to an increase of $2$-$5\%$ in F1 and $4$-$5\%$ in MA, and adding the proxy feature regeneration step on the same datasets leads to a further increase of $2$-$7\%$ in F1 and $5$-$7\%$ in MA.}}
\label{tab:ablation}
\end{table*}
\subsection{Evaluation Metrics and Methods}
\label{subsec:classification}
We use two standard metrics, F1 scores and mean classification accuracies (MAs), to evaluate all the methods. However, some prior methods have not reported MA, while others have not reported F1 scores. We, therefore, leave out the corresponding numbers in our evaluation as well and compare the methods with only the available numbers.
For the IEMOCAP dataset, we compare our accuracies with the following SOTA methods. 
\begin{enumerate}[noitemsep]
    \item \textbf{Yoon et al. \shortcite{DL1} }use only two modalities of the IEMOCAP dataset, text and speech, using an attention mechanism that learns to aligns the relevant text with the audio signal instead of explicitly combining outputs from the two modalities separately. The framework uses two Bi-linear LSTM networks. 
    \item \textbf{Kim et al. \shortcite{DL2} }focus on feature selection parts and hence use DBNs which they claim are better equipped at learning high-order non-linear relationships. They empirically show that non-linear relationships help in emotion recognition. 
    \item \textbf{Majumdar et al. \shortcite{DL3} }recognize the need of a more explainable and intuitive method for fusing different modalities. They propose a hierarchical fusion that learns bimodal and trimodal correlations for data fusion using deep neural networks. 
\end{enumerate}
For the CMU-MOSEI dataset, we compare our F1 scores with the following SOTA methods. 
\begin{enumerate}[noitemsep]
    \item \textbf{Zadeh et al. \shortcite{DL4} }propose a Dynamic Fusion Graph~(DFG) for fusing the modalities. The DFG can model n-modal interactions with an efficient number of parameters. It can also dynamically alter its structure and choose a fusion graph based on the importance of each n-modal dynamics. They claim that this is more interpretable fusion as opposed to the naive late fusion techniques.  
    \item \textbf{Choi et al. \shortcite{DL5} }use the text and speech modality of the CMU-MOSEI dataset. They extract feature vectors for text and speech spectrograms using Convolutional Neural Networks~(CNNs) architectures. They then use a trainable attention mechanism to leaner non-linear dependence between the two modalities. 
    \item \textbf{Sahay et al. \shortcite{DL6} }propose a tensor fusion network that explicitly models n-modal inter-modal interactions using an n-fold Cartesian product from modality embeddings.
\end{enumerate}
\begin{figure*}[h!]
    \centering
    \scalebox{0.9}{\includegraphics[width=\linewidth]{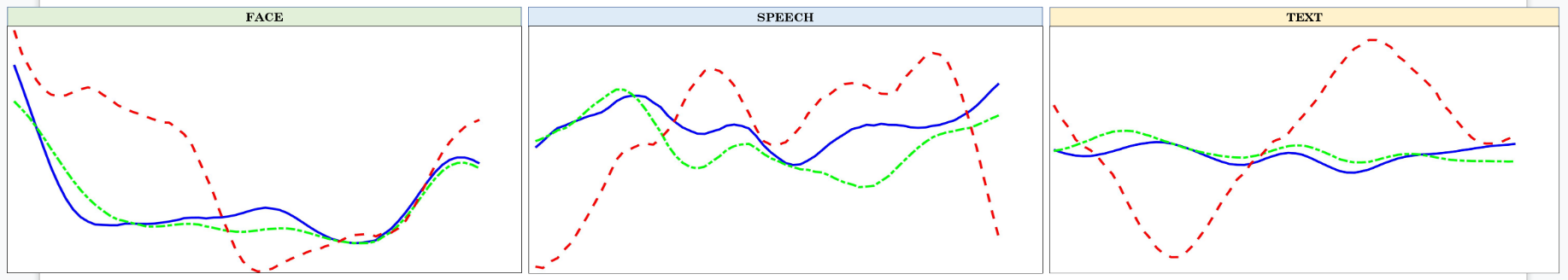}}
    \caption{\small{\textbf{Regenerated Proxy Feature Vector: }We show the quality of the regenerated proxy feature vectors for each of the three modalities. For the three graphs, we demonstrate the original feature vector~(blue), the ineffectual version of the modality because of added white Gaussian noise~(red) and the regenerated feature vector~(green). The mean $L_2$ norm distance between the original and the regenerated vector for the speech, text and face modality are all around $0.01\%$ of the $L_2$ norm of the respective data.}}
    \label{fig:proxyfeature}
\end{figure*}
\subsection{Analysis}
\label{subsec:analysis}
\noindent\textbf{Comparison with SOTA: }Evaluation of F1 scores and MAs of all the methods is summarized in Table~\ref{tab:accuracy}. We observe an improvement of $1$-$23\%$ in F1 scores and $5$-$10\%$ in MAs when using our method. \\\\
\noindent\textbf{Confusion Matrix: }We also show the confusion matrix (Figure \ref{fig:confusionmatrix}) to analyze the per-class performance of M3ER on IEMOCAP and CMU-MOSEI. We observe that more than $73$\% of the samples per class were correctly classified by M3ER. We see no confusions~($0$\%) between some emotion labels in the two confusion matrices, for instance `sad' and `happy' in IEMOCAP and `fear' and `surprise' in CMU-MOSEI.  Interestingly, we see a small set of data points getting confused between `happy' and `angry' labels for both datasets. We reason that this is because, in both situations, people often tend to exaggerate their cues. \\\\
\noindent\textbf{Interpretability of Multiplicative Layer: }We ran experiments to see the change of weights per sample point for each modality at the time of fusion to validate the importance of multiplicative fusion. Averaged over all the data points in the test set, when we corrupted the face modality, the average weight for the face modality dropped by 12\%, which was distributed tot he other modalities, text and speech. This is expected of the multiplicative layer, i.e. to adjust weights for each modality depending on the quality of the inputs.\\\\
\noindent\textbf{Qualitative Results: }Additionally, we show one sample per class from the CMU-MOSEI and IEMOCAP dataset that were correctly classified by M3ER in Figure~\ref{fig:qualitative} and Figure~\ref{fig:qualitative1}. \\\\
\noindent\textbf{Failure Case: } We also qualitatively show a data point in Figure~\ref{fig:failure} where M3ER fails to classify correctly. We observe that exaggerations of facial expressions and speech have led to a `happy' sample being classified by our model as `angry', a pattern also observed from the confusion matrices.\\
\subsection{Ablation Experiments}
\label{subsec:ablation}
\noindent\textbf{Original vs M3ER Multiplicative Fusion Loss.}
We first compare the original multiplicative fusion loss~\cite{multiplicative} (Equation~\ref{eq:original_loss}) with our modified loss (Equation~\ref{eq:modified_loss} on both IEMOCAP and CMU-MOSEI. As shown in Table~\ref{tab:ablation}, using our modified loss results in an improvement of $6$-$7\%$ in both F1 score and MA.\\
Next, to motivate the necessity of checking the quality of signals from all the modalities and implementing corrective measures in the case of ineffectual features, we corrupt the datasets by adding white Gaussian noise with a signal-to-noise ratio of $0.01$ to at least one modality in approximately $75$\% of the samples in the datasets. We then compare the performance of the various ablated versions of M3ER as summarized in Table~\ref{tab:ablation} and detailed below.\\\\
\noindent\textbf{M3ER -- Modality Check Step -- Proxy Feature Vector.} This version simply applies the multiplicative fusion with the modified loss on the datasets. We show that this results in a drop of $4$-$12\%$ in the overall F1 score and $9$-$12\%$ in the overall MA from the non-ablated version of M3ER.\\\\
\noindent\textbf{M3ER -- Proxy Feature Vector.} In this version, we perform the modality check step to filter out the ineffectual modality signals. This results in an improvement of $2$-$5\%$ in the overall F1 score and $4$-$5\%$ in the overall MA from the previous version. However, we do not replace the filtered out modalities with generated proxy features, thus having fewer modalities to work with. This results in a drop of $2$-$7\%$ in the overall F1 score and $5$-$7\%$ in the overall MA from the non-ablated version of M3ER.

Finally, with all the components of M3ER in place, we achieve an overall F1 score of $0.761$ on IEMOCAP and $0.856$ on CMU-MOSEI, and an overall MA of $78.2\%$ on IEMOCAP and $85.0\%$ on CMU-MOSEI. Additionally, we also show in Figure~\ref{fig:proxyfeature} that the mean $L_2$ norm distance between the proxy feature vectors regenerated by M3ER in and the ground truth data is around $0.01\%$ of the $L_2$ norm of the respective data.
\section{Conclusion, Limitations, and Future Work}
We present M3ER, a multimodal emotion recognition model that uses a multiplicative fusion layer. M3ER is robust to sensor because of a modality check step that distinguishes between good and bad signals to regenerate a proxy feature vector for bad signals. We use multiplicative fusion to decide on a per-sample basis which modality should be relied on more for making a prediction. Currently, we have applied our results to databases with three input modalities, namely face, speech, and text. 
Our model has limitations and often confuses between certain class labels. Further, we currently perform binary classification per class; however, human perception is rather subjective in nature and would resemble a probability distribution over these discrete emotions. Thus, it would be useful to consider multi-class classification in the future.
As part of future work, we would also explore more elaborate fusion techniques that can help improve the accuracies. We would like to extend M3ER for more than three modalities. As suggested in psychological studies, we would like to explore more naturalistic modalities like walking styles and even contextual information. 
\section{Acknowledgements}
This research is supported in part by ARO grant W911NF-18-1-0313. We would also like to thank Abhishek Bassan and Ishita Verma for initial few discussions on this project.

{\small
\bibliographystyle{aaai}
\bibliography{AAAI20-MittalT.4411.bib}
}

\end{document}